# Shuffle Instances-based Vision Transformer for Pancreatic Cancer ROSE Image Classification

Tianyi Zhang[1], Youdan Feng[1], Yunlu Feng[2], Yu Zhao[3], Yanli Lei[1], Nan Ying[1], Zhiling Yan[4], Yufang He[1], Guanglei Zhang[1]*

[1] Beijing Advanced Innovation Center for Biomedical Engineering, School of Biological Science and Medical Engineering, Beihang University, Beijing, China
[2] Department of Gastroenterology, Peking Union Medical College Hospital, Beijing, China
[3] Department of Pathology, Peking Union Medical College Hospital, Beijing, China
[4] School of Mathamatics, Sun Yat-sen University, Guangzhou, 510275, China
guangleizhang@buaa.edu.cn

**Abstract**

The rapid on-site evaluation (ROSE) technique can significantly accelerate the diagnosis of pancreatic cancer by immediately analyzing the fast-stained cytopathological images. Computer-aided diagnosis (CAD) can potentially address the shortage of pathologists in ROSE. However, the cancerous patterns vary significantly between different samples, making the CAD task extremely challenging. Besides, the ROSE images have complicated perturbations regarding color distribution, brightness, and contrast due to different staining qualities and various acquisition device types. To address these challenges, we proposed a shuffle instances-based Vision Transformer (SI-ViT) approach, which can reduce the perturbations and enhance the modeling among the instances. With the regrouped bags of shuffle instances and their bag-level soft labels, the approach utilizes a regression head to make the model focus on the cells rather than various perturbations. Simultaneously, combined with a classification head, the model can effectively identify the general distributive patterns among different instances. The results demonstrate significant improvements in the classification accuracy with more accurate attention regions, indicating that the diverse patterns of ROSE images are effectively extracted, and the complicated perturbations are significantly reduced. It also suggests that the SI-ViT has excellent potential in analyzing cytopathological images. The code and experimental results are available at https://github.com/sagizty/MIL-SI.

## Introduction

Pancreatic cancer is one of the most malignant tumors with the worst prognosis, whose 5-year survival rate is only 10% (Siegel et al. 2021, Zeng et al. 2018). Early diagnosis and treatment of pancreatic cancer can significantly improve the survival rate of patients. Endoscopic ultrasonography-guided fine-needle aspiration (EUS-FNA) is an important method for the pathological diagnosis of pancreatic cancer (Facciorusso et al. 2018, Matsubayashi et al. 2018). Combined with EUS-FNA surgery, rapid on-site evaluation (ROSE) can quickly obtain cytopathological images to improve diagnostic efficiency and reduce the number of punctures during surgery, reducing pain and the risk of postoperative infection (Yang, Liu and Sun 2019). However, the broader application of the ROSE technique is limited by the shortage of pathologists, which calls for on-site artificial intelligence (AI) to aid the workforce (van Riet et al. 2016). Computer-aided diagnosis (CAD) system shows high potential in solving this limitation (Calderaro and Kather 2021, Landau and Pantanowitz 2019, Marya et al. 2021).

At present, the research on CAD application for ROSE image classification is still in its infancy. In 2017, a multi-layer perceptron (MLP) model with an accuracy of 83.9% was established, based on the features extracted from the region of interest (ROI), such as cell contour and image intensity (Momeni-Boroujeni, Yousefi and Somma 2017). In 2018, Hashimoto et al. (2018) firstly applied deep learning to analyze ROSE images without manual feature extraction. This work trained and tested a model with 450 pathological images; the achieved sensitivity, specificity, and accuracy were all 80%. In 2020, the same group applied another deep learning strategy to the ROSE image classification and achieved 93% accuracy with 1440 images (Hashimoto 2020). The studies proved that deep learning has significant clinical value in cytopathological image analysis. However, the current works on ROSE image classification have paid little attention to the perturbation characteristics of images, and the interpretability can be further improved.

There are many challenges in the classification of ROSE images. Due to the influence of staining quality, the images often contain impurities, such as broken cells, pollution, fibers, light spots, and so on, which affect the classification performance. The difference in acquisition equipment also affects the image, creating differences in brightness, contrast, saturation, and resolution. The thickness of the slices brings difficulties in focusing, which means parts of the image are often vague, and the model should avoid regional misleading (Pantanowitz and Bui 2020). In addition to the perturbations of staining and image acquisition, the cytopathological characteristics of ROSE are also complex, making it difficult for the model to obtain a clear decision boundary in various images. With clinical knowledge, the differences between normal cells and cancer cells arise from local features (cell size, nucleocytoplasmic ratio, cell shape, nuclear membrane shape, chromatin distribution, etc.) and global features (overall orientation of cells, orientation of cell clusters, and spacing between cells, etc.). According to the gradient-weighted classification activation mapping (Grad-CAM), the existing models often focus too much on one of these features, which may lead to classification errors. Therefore, we need to find a method that can not only overcome the background noise of the image but also balance the global and local features.

The idea of multiple instance learning (MIL) brings us inspiration. MIL enables the model to focus on the differences among instances by bagging multiple instances and providing bag-level labels, showing potential in the classification task with complex background (Butke et al. 2021, Li et al. 2021, Yu et al. 2021). Inspired by MIL, we proposed a novel approach to enhance the backbone with a shuffle and re-grouping strategy. Keeping the instances with different staining and acquisition qualities in the same bag, the model focuses on the differences among cells rather than various perturbations of ROSE images under the supervision of bag-level soft labels. In addition, a combination of regression (REG) head and classification (CLS) head is designed to balance the feature extraction on the local and global features respectively. Combining with an alternative training strategy, we propose a shuffle instances-based Vision Transformer (SI-ViT) approach to classify the pancreatic ROSE images.

## Method

Towards the characteristics of the ROSE images, the SI-ViT approach is composed of two soft-label distributors, a Vision Transformer-based backbone, and two task-based heads (a REG head for soft label regression and a CLS head for classification). Two steps of the shuffle (SF) step and un-shuffle (USF) step are specially designed. In the SF step, instances from different images are shuffled and regrouped into bags, and after the feature extraction of the backbone, the REG head is used to regress the soft label of the re-grouped image. In the USF step, with un-shuffled images sent as input, the same REG head and an additional CLS head are used to predict the categories. The approach is described in **Fig. 1**.

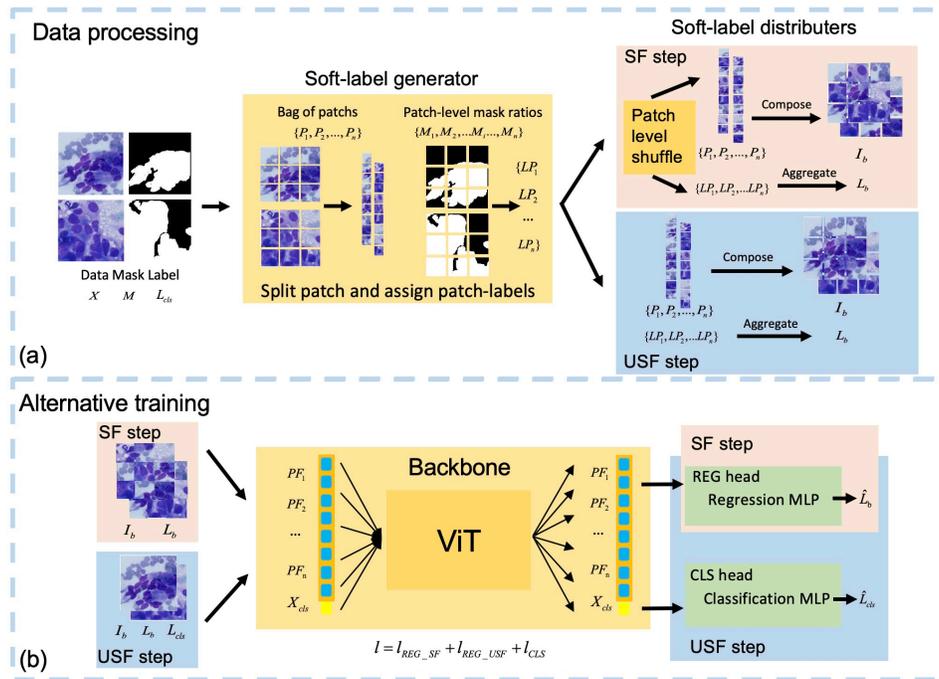

Figure 1: Overview of SI-ViT composed of two steps: SF and USF. In the data processing as illustrated in (a), the images are transformed into patches, and the patch label is calculated based on the corresponding masks. In the SF step, the instances are shuffled through the shuffle of patches from different bags, while the bags of instances remain unchanged in the USF step. Lastly, the bags are composed of images with their soft label aggregated from the patch-level label. In the proposed two-step training process in (b), the backbone alternatively extracts the features of shuffled and un-shuffled images. Then, the patch tokens are used to regress the bag-level soft label in the REG head. In the USF step, an additional CLS head is used to predict the categories of the input images based on the class token.

## Multiple Instance Learning

The concept of bags and instances in the SI-ViT approach comes from MIL. In the pancreatic cancer ROSE image classification task, each image is a bag, and cells in the image are instances. Defined by Dietterich et al. (1997), MIL models a bag of instances together, and the bag label is related to the instance label. The formulation of the bag label is

$$L = \begin{cases} 0, & if \ \sum L_i = 0 \\ 1, & otherwise \end{cases} \quad (1)$$

where $L$ is the bag label and $L_i$ is the label of instance $x_i$, $L, L_i \in [0,1]$, $i = 1, 2, ..., n$.

In many related works (Chikontwe et al. 2020, Oner et al. 2021, Shao et al. 2021), the MIL model can predict the bag label while the instance label is not accessible, and it performs better at learning the differences among instances. In addition to the prediction of the discrete label, the bag-level tumor purity can be regressed by MLP, showing that the regression MLP is applicable in the lightweight supervision on the whole bag to regress continuous label of distributions (Oner et al. 2021).

## Preprocessing and the Soft-label Distributors

**Preprocessing** After augmentation of the ROSE images and their masks, each ROSE image $X$ and its mask $M$ are sequentially divided into patches $\{P_1, P_2, ..., P_n\}$ and mask patches $\{M_1, M_2, ..., M_n\}$ by soft-label generator. The patch size is critical as the scale may influence the cells in each patch. In the shuffling process, the cells will be cut into pieces if the patch size is less than 16 pixels, and therefore the spatial features may be lost. On the contrary, if the patch size is too large, the cell groups may be reserved, causing the shuffling operation to be less effective in reducing the perturbations of images.

**Soft-label Distributors** After the patch split operation, each patch $P_i$ of the image $X$ is assigned with a patch label to represent its masked area of categories. The patch label is defined as

$$LP_i = \{MR_i, MR_{1i}, MR_{2i}, ..., MR_{Ki}\} \quad (2)$$

where $LP_i$ is the label of patch $P_i$, $MR_i$ is the ratio of masked pixels in $P_i$, and $MR_{ji}$ is the ratio of masked pixels of category $j$, $j = 1, 2, ..., K$. $K$ is the number of categories. For each patch, $MR_{ji}$ is defined as

$$MR_{ji} = \begin{cases} MR_i, & if \ j = k \\ 0, & otherwise \end{cases} \quad (3)$$

where $k$ is the category of $P_i$.

Then, adapted to two steps, we designed two soft-label distributors: SF distributor and USF distributor. In the SF soft-label distributor, the patches of bags in a batch are randomly shuffled to obtain mixed bags, while the bags remain un-shuffled in the USF soft-label distributor. Lastly, both distributors compose the bag of patches into an image $I_b$ and calculate its bag-level soft label $L_b$. To create pre-knowledge supervision for the regression task, the bag-level soft label is given as

$$L_b = \{\sum MR_i, \sum MR_{1i}, \sum MR_{2i}, ..., \sum MR_{Ki}\}. \quad (4)$$

The differences in bag labels between MIL and SI-ViT are illustrated in **Fig. 2**.

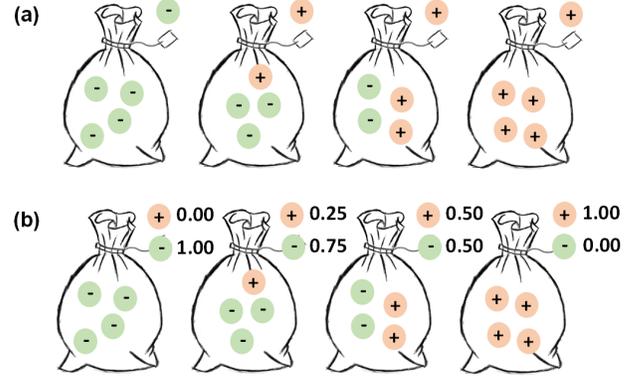

Figure 2: Bag Label of MIL (a) and SI-ViT (b)

## Vision Transformer Backbone

The Vision Transformer (ViT) (Dosovitskiy et al. 2020, Liu et al. 2021) has achieved inspiring results on various computer vision tasks, whose patch-based learning strategy is suitable for the SI-ViT approach, so we adopt ViT as the backbone. The feature sequence after feature extraction of the backbone is

$$\{X_{cls}, PF_1, PF_2, ..., PF_n\} = backbone(Embedding(I_b)) \quad (5)$$

where $X_{cls} \in R^D, PF_i \in R^D$, $D$ is the embedding dimension, $X_{cls}$ is the CLS token, and $PF_i$ is the output token of backbone.

After the feature extraction, the CLS token $X_{cls}$ is used for the CLS head to predict the categories, and the rest of tokens $\{PF_1, PF_2, ..., PF_n\}$ representing the patch features are reserved for the REG head.

Therefore, the predictions of bag-level soft label $\hat{L}_b$ and classification label $\hat{L}_{cls}$ are defined as

$$\hat{L}_b = REG(\{PF_1, PF_2, ..., PF_n\}) \quad (6)$$

$$\hat{L}_{cls} = CLS(X_{cls}). \quad (7)$$

## Two Task-based Heads and Dataflow

Adapting to the two labels, the SI-ViT model is designed with two MLP heads: a CLS head and a REG head, which predict the classification label and regress the bag-level soft label simultaneously. To our knowledge, this is the first work using the patch-level shuffle strategy on ViT to match the characteristics of pathological image classification.

The dataflow of the SI-ViT is illustrated in **Fig. 1**. For each ROSE image considered as a bag within a batch, the image and its mask are divided into patches, and the soft label of each patch is assigned based on its mask. Secondly, two dataflows are specially designed. In the SF step, the bags within a batch are shuffled on patch level randomly, so the cell instances from different patches are grouped in a new bag. The global features are randomized through the shuffle strategy, so the model learns local features by learning the differences among instances. On the contrary, the bags are not shuffled in the USF step to maintain their global features. After the bag distribution, each bag of patches is regrouped into an image with the corresponding bag-level soft label as input for the model.

Then, the model is trained in two steps, and both steps use the backbone to extract features. In the SF step, the features extracted from the shuffled images are modeled in the REG head to regress the bag-level soft label, which can make the model specially model the cross-instance distributions. In the USF step, the REG head is also used based on un-shuffled images, and the CLS head is used to predict the classification label from the CLS token in the Transformer. Lastly, the loss is calculated by

$$l = l_{REG\_SF} + l_{REG\_USF} + l_{CLS} \tag{8}$$

where $l_{REG\_SF}$ is loss from REG head in the SF step, $l_{REG\_USF}$ is loss from REG head in the USF step and $l_{CLS}$ is loss from CLS head in the USF step.

## Experiments and Results

### Dataset and Experimental Setting

The EUS-FNA and ROSE diagnoses were performed in Peking Union Medical College Hospital, and the ROSE images were collected under the supervision of senior pathologists. A total of 1773 pancreatic cancer images and 3315 normal pancreatic cell images were collected, and senior pathologists confirmed the classification labels and segmentation results. An open-source dataset was compared to further prove the efficiency of SI-ViT on histopathological samples. The p-RCC subtyping dataset contains 62 type 1 and 109 type 2 cases of diagnostic whole slide images (WSI) from 171 patients. These two datasets were divided into training, validation, and test sets at the ratio of 7:1:2.

The same data augmentation strategy was designed in each experiment to deal with the data scarcity and image perturbations. The model was trained for 50 epochs, and the model with the highest validation accuracy was saved as the output model. The same training parameters or higher-performed settings were used in all experiments. To compare the performance of our proposed SI-ViT, several widely used CNNs and Transformers were compared as counterpart models. Furthermore, to prove the effectiveness of the shuffle instances strategy, the USF-ViT model using only the USF step (with the REG head working on the un-shuffled bag) and the backbone model (ViT) were compared. More details of datasets and experimental settings are provided in the supplementary material.

The experimental results of SI-ViT and its counterparts were measured by accuracy, precision, recall, specificity, and F1-score. The Grad-CAM was used to visualize the attention regions to reveal the interpretability of the models.

### Comparison with SOTAs

SI-ViT aims to enhance a patch-based backbone with little changes within the model and a simple but effective design of training strategy. To verify the effectiveness of our approach, we compared SI-ViT with other models under the same experimental conditions. As a promising option in the cytopathology research, two SOTAs with patch-based learning: basic ViT and Swin Transformer; and seven widely applied state-of-the-art CNNs including VGG-16, VGG-19, EfficientNet-b3, ResNet50, Inception-V3, Xception and MobileNet-V3, and two recent hybrid methods Conformer and Cross-former are compared.

As shown in **Table 1**, SI-ViT achieves the best results among these models in accuracy, precision, recall, etc. Compared with other models, the accuracy of SI-ViT is improved by 3.2%-4.8%. Other indicators, especially precision and recall, are improved by approximately 6%. In the analysis of ROSE images, the results of SI-ViT are significantly better than the earlier work (Hashimoto et al. 2018, Hashimoto 2020) (89% and 93% accuracy), and SI-ViT is fully validated on a much larger dataset of 5088 images compared with their 1440 images.

| Models | Accuracy (%) | Precision (%) | Recall (%) | F1-score (%) |
|---|---|---|---|---|
| VGG-16 | 90.65 | 86.27 | 87.01 | 86.64 |
| VGG-19 | 90.06 | 90.42 | 79.94 | 84.86 |
| Mobilenet-V3 | 89.57 | 91.06 | 77.68 | 83.84 |
| EfficientNet-b3 | 89.57 | 85.03 | 85.03 | 85.03 |
| Inception-V3 | 90.75 | 86.72 | 86.72 | 86.72 |
| Xception | 90.94 | 91.46 | 81.64 | 86.27 |
| ResNet50 | 90.75 | 87.36 | 85.88 | 86.61 |
| Swin Transformer | 89.17 | 86.75 | 81.36 | 83.97 |
| ViT | 91.24 | 89.55 | 84.75 | 87.08 |
| Conformer | 89.67 | 90.82 | 78.25 | 84.07 |
| Cross-former | 89.67 | 86.94 | 82.77 | 84.80 |
| **SI-ViT** | **94.00** | **91.98** | **90.68** | **91.32** |

Table 1: Comparison with SOTAs on ROSE

In addition, by visualizing the Grad-CAM, our model obtains the most accurate attention regions compared with other models (**Fig. 3**). Both in positive and negative images, our model can accurately focus on the areas of the cell clusters instead of backgrounds, presenting solid interpretability in the cell identification. In most cases, by shuffle instances strategy, the SI-ViT extracts spatial features of each cell instance and models their relations globally. As further proof in **Fig. 4**, by visualizing its attention on shuffled bags, the SI-ViT can identify all the normal and cancer cells and shows no redundant attention spent on the background.

The SI-ViT considers the positive samples to negative in most misclassified samples. The problem of misclassification is given by taking two examples, as shown in **Fig. 5**. By the analysis of the senior pathologists, SI-ViT presents reasonable attention regions on the cancerous regions, although the activation values are not ideal. However, both samples have misleading distributions as the normal cells; furthermore, the fluctuation of the squeezed cells in these samples may also mislead junior doctors.

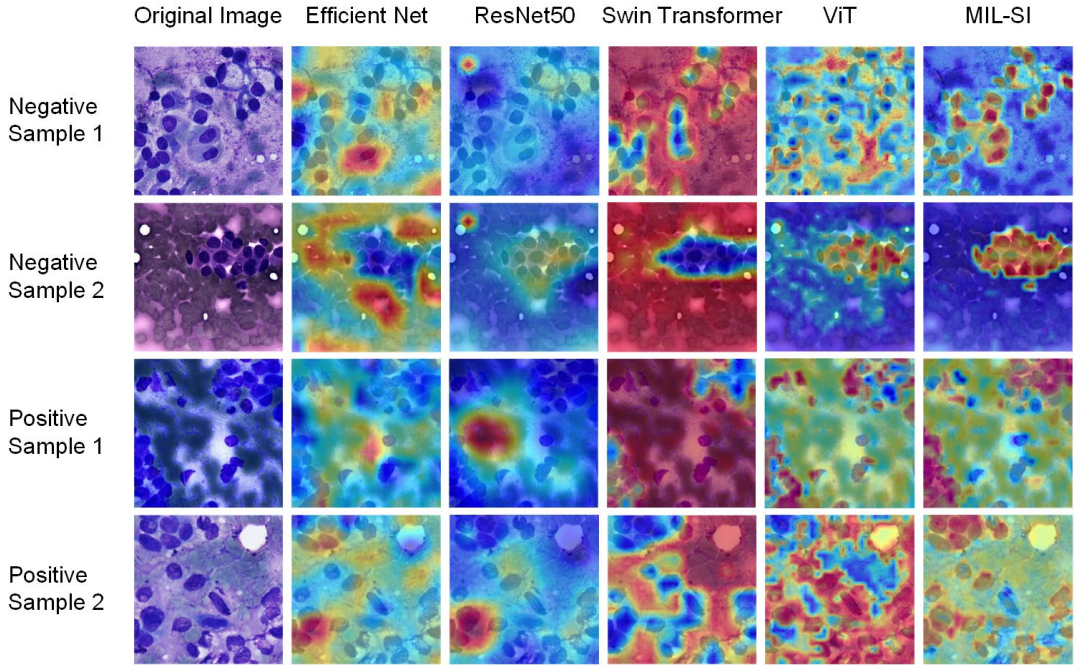

Figure 3: Attention Regions of Typical Samples in SI-ViT

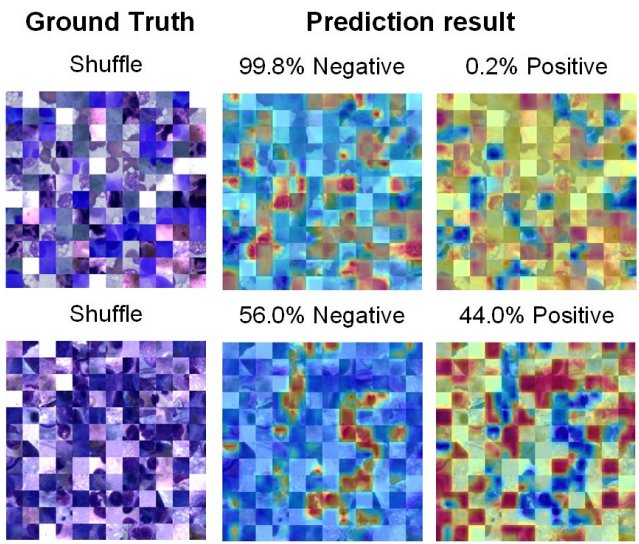

Figure 4: Grad-CAM of Shuffle Samples in SI-ViT

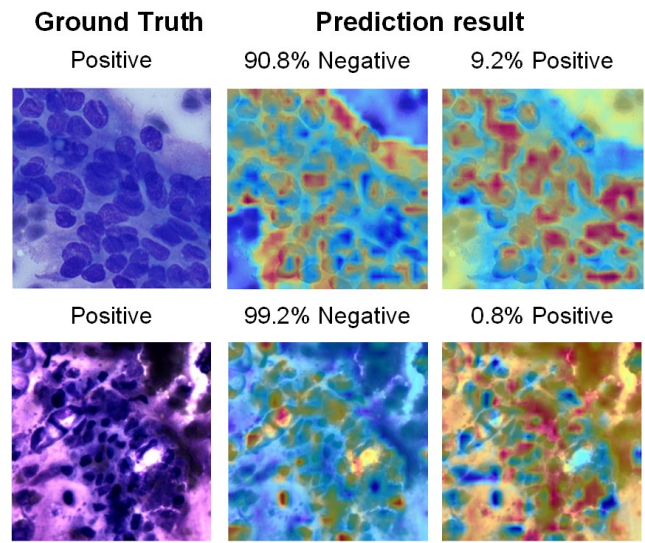

Figure 5: Grad-CAM of Failure Samples in SI-ViT

## Comparison with Data Augmentations

The SI-ViT is proposed with a dataflow-drive strategy to train the backbone with alternative regrouped images and un-shuffled images. The idea of altering the input of the backbone meets the essence of data augmentations. Therefore, three wildly applied data augmentation methods: CutOut, MixUp, and CutMix, are compared to prove the effectiveness of the shuffle strategy. To be specific, the CutOut randomly covers an image; the MixUp randomly joins two images and gives a weight ratio as the soft label; the CutMix pastes a random part of another image to the target image and uses the image occupation ratio as the soft label.

In **Table 2**, three data-augmentation methods achieve satisfying results compared with naive ViT. The SI-ViT achieves the highest results, proving the idea of shuffle instances can enhance the basic model most.

| Model | Strategy | Accuracy (%) | Precision (%) | Recall (%) | F1-score (%) |
|---|---|---|---|---|---|
| ViT | Naive | 91.24 | 89.55 | 84.75 | 87.08 |
| ViT | CutOut | 92.32 | 91.07 | 86.44 | 88.70 |
| ViT | MixUp | 92.52 | 88.83 | 89.83 | 89.33 |
| ViT | CutMix | 92.72 | 89.55 | 89.55 | 89.55 |
| ViT | **SI-ViT** | **94.00** | **91.98** | **90.68** | **91.32** |

Table 2: Comparison with Data Augmentation on ROSE

## Comparison on the p-RCC Dataset

The primary experiments are also applied to an open-source dataset (p-RCC) to evaluate the SI-ViT. The SI-ViT on the ROSE dataset utilizes a patch size of 32, while in p-RCC is 128 as the scale of the instance in histopathology (tissues) differs from the cytopathological images (cells).

| Model | Strategy | Accuracy (%) | Precision (%) | Recall (%) | F1-score (%) |
|---|---|---|---|---|---|
| VGG-16 | Naive | 92.23 | 93.07 | 86.24 | 89.52 |
| VGG-19 | Naive | 93.64 | 95.05 | 88.07 | 91.43 |
| Mobilenet-V3 | Naive | 80.92 | 76.70 | 72.48 | 74.53 |
| EfficientNet-b3 | Naive | 79.15 | 72.32 | 74.31 | 73.30 |
| Inception-V3 | Naive | 87.99 | 83.19 | 86.24 | 84.68 |
| Xception | Naive | 88.69 | 96.39 | 73.39 | 83.33 |
| ResNet50 | Naive | 88.34 | 90.43 | 77.98 | 83.74 |
| Conformer | Naive | 92.93 | 96.84 | 84.40 | 90.20 |
| Cross-former | Naive | 95.05 | 96.12 | 90.83 | 93.40 |
| Swin Transformer | Naive | 94.70 | 93.52 | 92.66 | 93.09 |
| ViT | Naive | 92.58 | 94.00 | 86.24 | 89.95 |
| ViT | CutMix | 93.29 | 95.92 | 86.24 | 90.82 |
| ViT | CutOut | 93.99 | 90.35 | 94.50 | 92.38 |
| ViT | MixUp | 93.29 | 92.45 | 89.91 | 91.16 |
| **SI-ViT** | **SI-ViT** | **95.76** | **98.02** | **90.83** | **94.29** |

Table 3: Results of the Counterpart Methods on P-RCC

As shown in Table 3, compared with SOTAs and other strategies, the results prove the significant increase in SI-ViT. Furthermore, compared with the recent work by Gao et al. (2021), the SI-ViT achieves a significant 2.75% accuracy increase (95.76% vs. 93.01%). The experiments on the p-RCC dataset prove the generalizability of SI-ViT, as the idea of shuffle and regrouping instances can be used not only in cell instances but also in tissue instances.

## Ablation Studies

To evaluate the structural improvement and explore the efficiency of the SI-ViT, a series of ablation studies were drawn out under the same training and validating setting.

**The Effectiveness of SI-ViT Structure and Shuffle Step**
To verify the effectiveness of the shuffle instances strategy, we conducted the USF-ViT model with both the REG head and CLS head to maintain the same structure and calculation but without the shuffle operation. As an ablation model of SI-ViT, USF-ViT uses the REG head to regress the soft label on un-shuffled samples and predict the categories. Without the REG head, the SI-ViT is naive ViT, illustrated in the first column with USF-ViT and SI-ViT.

| Model | Strategy | Accuracy (%) | Precision (%) | Recall (%) | F1-score (%) |
|---|---|---|---|---|---|
| ViT | No REG head & No shuffle | 91.24 | 89.55 | 87.08 | 87.08 |
| USF-ViT | No shuffle | 92.13 | 90.06 | 87.01 | 88.51 |
| SI-ViT | **SI-ViT** | **94.00** | **91.98** | **90.68** | **91.32** |

Table 4: Results of the Ablation Models on ROSE

In **Table 4**, introducing the soft label and REG head can improve the classification accuracy (91.24% to 92.13%) even without the shuffle operation. With the un-shuffled bag modeling process in the USF step, the introduction of the soft label with REG head brings a 0.89% increase in ACC, as the model simultaneously learns global features and distributions. But with the shuffle instances strategy, the augmented distributions of instances lead to a significant 2.76% increase in ACC. The SI-ViT achieved the highest results with the modeling process on the regrouped instances, indicating that the perturbations among samples can be reduced and the relations of pancreatic cells can be better modeled.

**The Selection of Backbone** As a patch-based approach, the essence of instance regrouping is evaluated on different backbones. Firstly, the ViT and Swin Transformer are used to evaluate the efficiency of SI-ViT on the patch-based backbones. Then the ResNet50 is specially applied to observe the ConvNets, which are not patch-based approaches. The REG head adjusts based on the output feature maps of these backbones. Lastly, the SI-ViT is evaluated with different scales of backbone ViT.

In **Table 5**, the proposed SI-ViT achieves the highest results with the backbone ViT-base. In the scaling experiments, the ViTs achieve better results except for ViT-tiny, in which the limited model size may be difficult to model both REG head and CLS head. Bigger ViTs may be overfitting on a limited dataset as the results increased less than the SI-ViT with ViT-base.

| Backbone | Strategy | Accuracy (%) | Precision (%) | Recall (%) | F1-score (%) |
|---|---|---|---|---|---|
| ViT-tiny | Naive | 91.93 | 90.00 | 86.44 | 88.18 |
| ViT-small | Naive | 90.35 | 88.32 | 83.33 | 85.76 |
| ViT-base | Naive | 91.24 | 89.55 | 84.75 | 87.08 |
| ViT-large | Naive | 90.94 | **92.53** | 80.51 | 86.10 |
| ViT-tiny | SI-ViT | 91.63 | 89.21 | 86.44 | 87.80 |
| ViT-small | SI-ViT | 91.83 | 88.83 | 87.57 | 88.19 |
| **ViT-base** | **SI-ViT** | **94.00** | 91.98 | **90.68** | **91.32** |
| ViT-large | SI-ViT | 92.81 | 91.45 | 87.57 | 89.47 |
| ResNet50 | SI-ViT | 80.71 | 92.47 | 48.59 | 63.70 |
| Swin Transformer | SI-ViT | 92.81 | **94.60** | 84.18 | 89.09 |
| **ViT-base** | **SI-ViT** | **94.00** | 91.98 | **90.68** | **91.32** |

Table 5: Results of the Different Backbones on ROSE

With the patch-based method Swin Transformer, SI-ViT achieves a significant increase compared with Naive Swin Transformer in **Table 1** (89.17% in ACC). However, the results decay when SI-ViT meets the non-patch-based method ResNet50 backbone. The results indicate that the shuffle method requires a patch modeling ability, which the non-patch-based methods may not naturally have.

**The Effectiveness of Image Size and Patch Size** As a patch-based approach, the scale of each regrouped patch affects the shuffle and regrouping object as they are assigned as the instances. Different scales of patches represent different patterns. In the ROSE dataset, the bigger patches may contain more distributive patterns of cell clusters, and the smaller ones contain more individual cell features.

| Image Size | Patch Size | Accuracy (%) | Precision (%) | Recall (%) | F1-score (%) |
|---|---|---|---|---|---|
| 224 | 16 | 91.54 | 91.10 | 83.90 | 87.35 |
| 224 | 32 | 93.31 | 90.17 | **90.68** | 90.42 |
| 384 | 16 | 93.60 | 92.88 | 88.42 | 90.59 |
| **384** | **32** | **94.00** | 91.98 | **90.68** | **91.32** |
| 384 | 64 | 93.11 | 91.04 | 88.98 | 90.00 |
| 384 | 128 | 92.62 | **92.92** | 85.31 | 88.95 |

Table 6: Results of the Different Sizes on ROSE

In **Table 6**, the proposed SI-ViT performs best when the shuffling patch contains approximately one cell. When a cell serves as an instance, the regrouping design can reduce the perturbation and model each cell instance upon its spatial features. The results indicate that the model shuffling at the cell scale performs best, and the visualization of attention regions proves the same in **Fig. 6**. More experiment results are provided in the supplementary material.

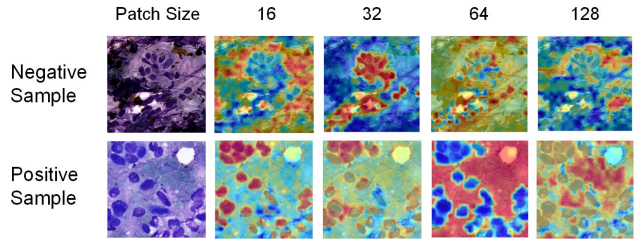

Figure 6: Attention Regions with Different Patch Sizes

## Conclusion

In conclusion, the proposed SI-ViT presents a novel view of the modeling process of pancreatic cancer ROSE images. As an instance-aware strategy, the shuffle and regrouping design in SI-ViT introduces the MIL idea within the dataflow. The design reveals the cell features as individuals and then models the relations among them. Cells' prominent spatial features and distributive patterns are balanced beyond the complicated perturbations during the sampling. With the solid evaluation of strategy design and inspiring experimental results, SI-ViT achieves the new SOTA results and promising interpretability. In general, the model focuses on cells, and the background perturbations are effectively reduced.

The two heads: the CLS head and the REG head, introduce only limited changes to the original structure, which enables the community to dive into a more profound view following the patch-based learning methodology. The straightforward loss design balances the cell instance identification and cross-instance modeling. The alternative training design and composed loss backward strategy make the SI-ViT robust and easy to deploy to other downstream tasks.

As the instance regrouping method applies to other datasets, the generalizability is evaluated with the p-RCC dataset. The instance scale is elastic, and the features may depend on cell clusters. As the modeling process on the ROSE dataset, the SI-ViT balances the identification of instances and cross instance modeling. The SOTA results with SI-ViT prove the effectiveness of this novel design. The exceptional view of instance relations sheds light on the potential modeling formulation.

The inspiring results and attention visualizations of ROSE images prove the implementation possibility of SI-ViT in hospitals. By deploying AI on-site system instead of a human pathologist on-site analysis, the diagnosis workflow can be significantly boosted, saving more time in the treatment process. To take a step forward, the border expansion of ROSE in-surgery diagnosis can benefit more clinical centers and reach more patients.

# Supplementary Material

## Dataset

The EUS-FNA and ROSE diagnoses were performed in an anonymous hospital, and the ROSE images were collected under the supervision of senior pathologists. The enrolled images were acquired by two microscope digital cameras (Olympus BX53 and Nikon Eclipse Ci-S) through the acquisition heads of Basler ScA1 and Olympus DP73. The images were saved in 'jpg' format with a resolution of 1390*1038. A total of 1773 pancreatic cancer images and 3315 normal pancreatic cell images were collected, and senior pathologists confirmed the classification labels and segmentation results. The 5088 images were divided into training, validation, and test sets at the ratio of 7:1:2.

An open-source dataset was compared to further prove the efficiency of SI-ViT on histopathological samples. The p-RCC subtyping dataset contains 62 type 1 and 109 type 2 cases of diagnostic whole slide images (WSI) from 171 patients (scanned at 40x). A total of 1162 (613 vs. 549) ROIs in 2000×2000 sizes were selected by two experienced pathologists, approximately 10 ROIs for every type 1 case and 5 for type 2. Following the same ratio of 7:1:2, the p-RCC dataset is randomly divided into training, validation, and test sets.

## Experimental Setting

The same data augmentation strategy was designed in each experiment to deal with the data scarcity and image perturbations. In the data augmentation process, the data and mask were randomly rotated, and the center area of 700*700 pixels was reserved. Secondly, the area was resized to 384*384 pixels, and the random vertical and horizontal flip were applied. Lastly, for each ROSE image, the PyTorch ColorJitter (with brightness = 0.15, contrast = 0.3, saturation = 0.3, and hue = 0.06) was used to recreate the perturbations during the sampling process. In the validation and test processes, only the center-cut and the resizing operations were applied, obtaining the center area of 700*700 pixels and resizing it to 384*384 pixels.

The model was trained for 50 epochs, and the model with the highest validation accuracy was saved as the output model. The Adam optimizer was applied with a learning rate of 1e-5 and a momentum of 0.05. The cosine learning rate decay strategy was adopted to reduce the learning rate twenty times sequentially in the training process. The same training parameters or higher-performed settings were used in all experiments.

To compare the performance of our proposed SI-ViT on the ROSE classification task, several widely used CNNs and Transformers were compared as counterpart models. Moreover, to prove the effectiveness of the shuffle instances strategy, the USF-ViT model using only the USF step (with the REG head working on the un-shuffled bag) and the backbone model (ViT) were compared. All models were built based on the timm library, and the transfer learning strategy was applied to all models with their official weights. The experiments were carried out and recorded online on the Google CoLab pro+ platform. As a lightweight approach, a single 16 GB Nvidia P100-PCIe GPU was used with Python version 3.7.12 and Pytorch version 1.10.0+cu111.

## The Effectiveness of Head Weight

As illustrated in Method, the SI-ViT has two heads: the REG head is used in both SF and USF steps, while the CLS head will is used in the USF step. As it is a bias-setting towards the instance identifying process (SF step) and group modeling (USF step for classification), the weight ratio of the heads (CLS: REG_USF: REG_SF) may affect the results.

| Weight Ratio | Accuracy (%) | Precision (%) | Recall (%) | F1-score (%) |
|---|---|---|---|---|
| 1.0: 0.5: 0.5 | 91.73 | 84.62 | **93.22** | 88.71 |
| 1.0: 1.2: 1.2 | 92.52 | 87.77 | 91.24 | 89.47 |
| 1.0: 1.5: 1.5 | 93.31 | 91.81 | 88.70 | 90.23 |
| 1.0: 1.8: 1.8 | 93.41 | 91.12 | 89.83 | 90.47 |
| 1.0: 2.5: 2.5 | 93.50 | **92.35** | 88.70 | 90.49 |
| 1.0: 3.0: 3.0 | 93.60 | 91.40 | 90.11 | 90.75 |
| **1.0: 1.0: 1.0** | **94.00** | 91.98 | 90.68 | **91.32** |

Table S1: Results of Different Head Weights

As shown in **Table S1**, the results indicate that the SI-ViT achieves higher performance when the weight bias of the CLS head, REG head in the USF step, and REG head in the SF step is naive 1:1:1. The findings indicate that the two heads should be treated equally in the alternative training process.

## A Further Discussion on Instance Size

As illustrated in **Table 6**, the scale of instances is vital in the shuffling process, as it links to the feature scale in the modeling process. The feature scale of different samples from different datasets varies, and the scale within the same dataset indicates different patterns. In the ROSE dataset, SI-ViT performs the highest test accuracy under the patch size of 32, while the p-RCC dataset has the highest result in 128. The Grad-CAM is used to analyze the modeling process when the SI-ViT faces the regrouped bags.

In **Fig. S1**, the SI-ViT with patch sizes 32, 64, and 128 all achieve high confidence in classifying cancer cells. However, the attention regions in the 64 and 128 are not ideal as some cancer cells are ignored. The cells are generally focused on the patch size of 16, but the confidence is much lower. The patch size of 32 achieves the best attention illustration among the counterparts. The visualizations show that the instance grouping strategy can have the model to model the relations of the cells. However, with a particular design, when the patch scale setting suits an instance most, the shuffling strategy achieves the highest results. The findings indicate that the essence of regrouping is to regroup the instances. A similar result can be found in the p-RCC dataset, where the instance is cell groups of a larger scale (**Fig. S2**).

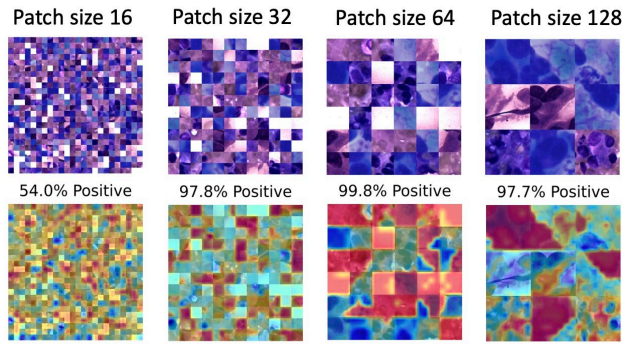

Figure S1: Attention Regions with Different Patch Sizes on ROSE

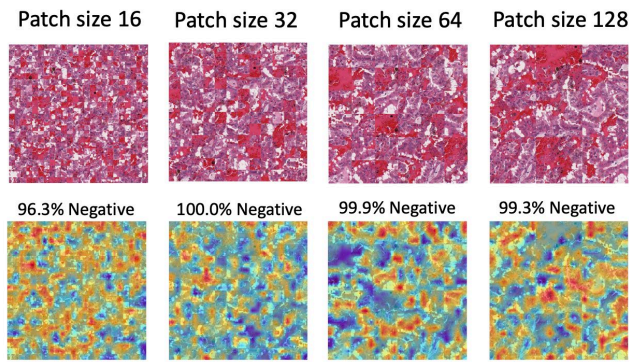

Figure S2: Attention Regions with Different Patch Sizes on p-RCC